\documentclass[newfonts=false,format=sigconf,10pt,letterpaper,table,xcdraw]{acmart}
\usepackage{amsmath}
\usepackage{listings}
\usepackage{setspace}
\usepackage{indentfirst}
\usepackage{graphicx}
\usepackage{subfigure}
\usepackage[justification=centering]{caption}
\usepackage{makecell}
\usepackage{url}
\usepackage{multirow}
\usepackage{xcolor}
\usepackage{diagbox}
\usepackage{pgfplotstable}
\usepackage{pgfplots}
\usepgfplotslibrary{groupplots}
\usetikzlibrary{patterns}
\usepackage{verbatim}
\usepackage{color}
\usepackage{changepage}
\usepackage{enumitem}

\makeatletter
\newif\if@restonecol
\makeatother

\usepackage[ruled,vlined]{algorithm2e}

\newcommand*\circled[1]{\tikz[baseline=(char.base)]{\node[shape=circle,draw,inner sep=0pt] (char) {#1};}}

\begin{document}

\title{\textsf{PAM}: When Overloaded, Push Your Neighbor Aside!}
\author{\vspace{-1\baselineskip}Zili Meng$^\dag$, Jun Bi$^\#$, Chen Sun$^\dag$, Shuhe Wang$^\dag$, Minhu Wang$^\dag$, Hongxin Hu$^\S$}
\affiliation{\vspace{-.5\baselineskip}$^{\dag\#}$Institute for Network Sciences and Cyberspace, Tsinghua University\\$^{\dag\#}$Beijing National Research Center for Information Science and Technology (BNRist)\\$^\dag$\{mengzl15, c-sun14, wang-sh14, wang-mh15\}@mails.tsinghua.edu.cn, $^\#$junbi@tsinghua.edu.cn \\$^\S$School of Computing, Clemson University, hongxih@clemson.edu}
\thanks{This poster is supported by the National Key R\&D Program of China (No.2017YFB0801701), the National Science Foundation of China (No.61472213), the Tsinghua Initiative Research Program (No.20181080342), and the National Training Program of Innovation and Entrepreneurship for Undergraduates. Jun Bi is the corresponding author.}
\vspace{-1\baselineskip}

\copyrightyear{2018} 
\acmYear{2018} 
\setcopyright{acmlicensed}
\acmConference[SIGCOMM Posters and Demos '18]{ACM SIGCOMM
2018 Conference Posters and Demos}{August 20--25, 2018}{Budapest,
Hungary}
% \acmBooktitle{SIGCOMM Posters and Demos '18: ACM SIGCOMM 2018 Conference Posters and Demos, August 20--25, 2018, Budapest, Hungary}
\acmPrice{15.00}
\acmDOI{10.1145/3234200.3234215}
\acmISBN{978-1-4503-5915-3/18/08}

\renewcommand{\shortauthors}{Zili Meng, \textit{et al}.}

\renewcommand{\authors}{Zili Meng, Jun Bi, Chen Sun, \textit{et al}.}

\fancyhead[LE,RO]{\textsf{SIGCOMM Posters and Demos '18}\\\textsf{Budapest, Hungary, August 20--25}}




\maketitle

\vspace{-1\baselineskip}
\section{Introduction}
\label{sec:intro}

Network Function Virtualization (NFV) enables efficient development and management of network functions (NFs) by replacing dedicated middleboxes with virtualized Network Functions (vNFs). When a vNF is overloaded, network operators can easily scale it out by creating a new vNF instance and balancing the load between two instances. Meanwhile, network operators usually require packets to be processed by multiple vNFs in a certain sequence, which is referred to as a service chain~\cite{kumar2015service}. However, the introduction of NFV results in \textit{high latency}. Virtualization techniques in NFV significantly increase processing latency~\cite{sigcomm2017nfp}. 
% Moreover, the latency of a service chain may grow linearly with the chain length (possibly \textit{eight} or longer~\cite{kumar2015service}). 
To address this problem, many research efforts from both industry~\cite{netronome-smartnic-cx10gb} and academic communities~\cite{socc2017uno} introduce programmable Network Interface Cards based on Network Processing Units (NPUs), \textit{i.e.} SmartNICs, to accelerate NFV. With the advantage of high performance and resource efficiency, offloading vNFs from CPU to SmartNIC brings significant performance benefits.

\begin{figure}
\centering
\subfigure[Service chain before migration.]{
	\centering
    \includegraphics[scale=0.38]{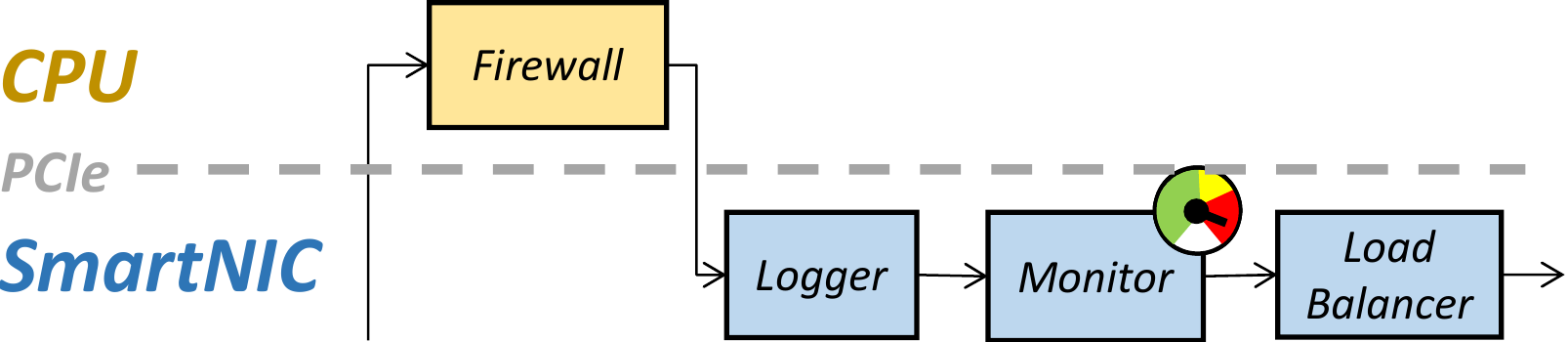}
    % \vspace{-5\baselineskip}
    \label{fig:migrate-origin}
    % \vspace{-.5\baselineskip}
}
\vspace{-.5\baselineskip}
\subfigure[Service chain after migration with the naive solution.]{
	\centering
    \includegraphics[scale=0.38]{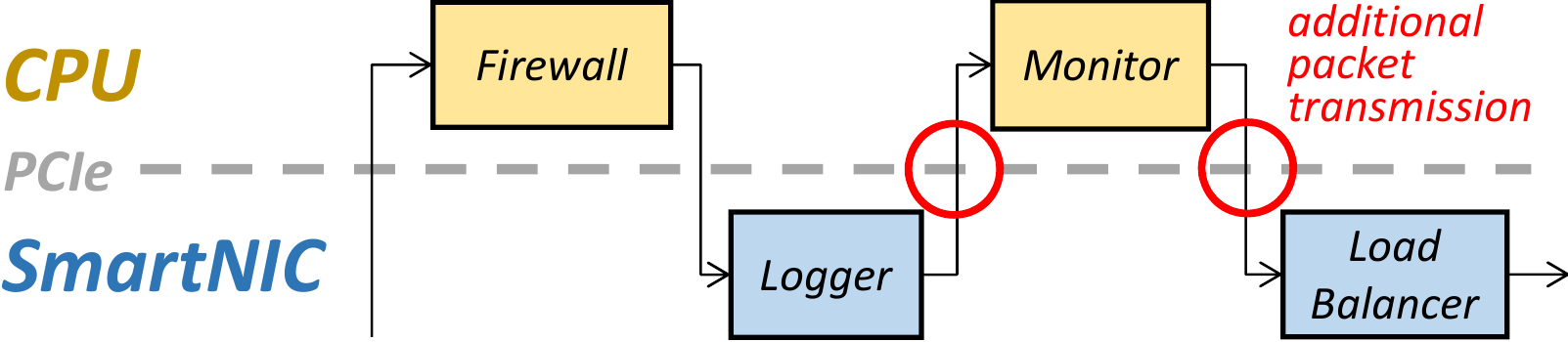}
    \label{fig:migrate-casual}
}
\vspace{-.5\baselineskip}
\subfigure[Service chain after migration with \textsf{PAM}.]{
	\centering
    \includegraphics[scale=0.38]{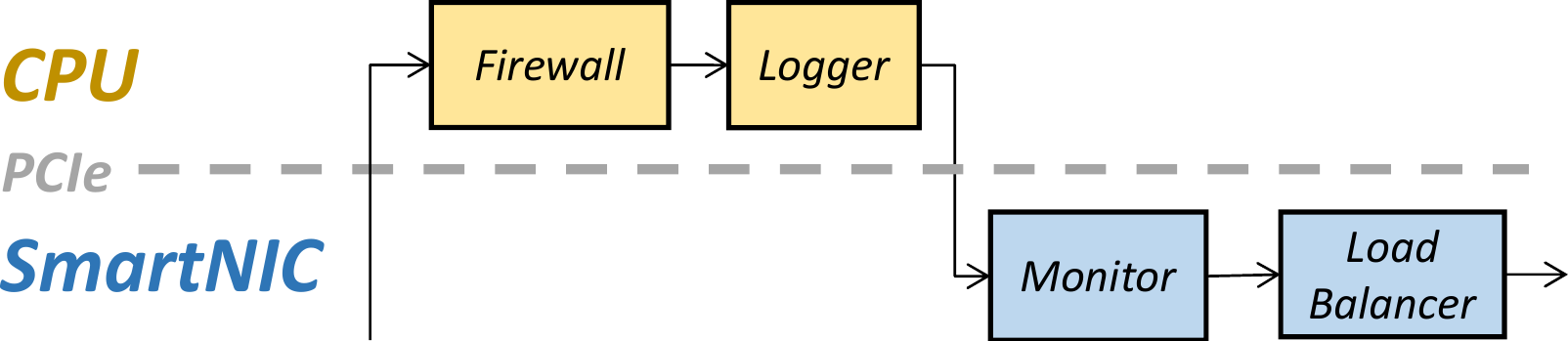}
    \label{fig:migrate-smartchain}
}
\vspace{-1\baselineskip}
\caption{Comparison of \textsf{PAM} with casual migration. The service chain is derived from~\cite{sigcomm2017nfp}.}
\label{fig:migrate}
\vspace{-1.5\baselineskip}
\end{figure}

Meanwhile, as the network traffic fluctuates, NFs on SmartNIC can also be overloaded~\cite{socc2017uno}. If we naively refer to the scaling out solutions for CPU, we have to introduce one more SmartNIC to alleviate the hot spot, which is hardly possible since each server is usually equipped with one or two SmartNICs only. UNO~\cite{socc2017uno} proposed to alleviate the overload situation by % and designed mechanisms for \textit{safe} and \textit{efficient} migration between SmartNIC and CPU~\cite{37gember2014opennf}. %However, an intuition is that deciding \textit{which vNFs to migrate} is also an important problem. 
identifying the \textit{bottleneck vNF} with minimum processing capacity and migrating it to CPU. However, this intuitive naive solution may increase the latency of the service chain. As shown in Figure~\ref{fig:migrate-casual}, if \textit{Monitor} is the bottleneck vNF and we migrate it to CPU, packets have to be transmitted over PCIe for two more times. This adds tens of microseconds latency according to our experiments, which may be unacceptable for latency-sensitive applications~\cite{sigcomm2017nfp}.

To address this problem, in this poster, we propose \textsf{PAM}, the Push Aside Migration scheme, which identifies the \textit{right vNFs} to migrate to alleviate the hot spot on SmartNIC without introducing long-term performance degradation. We consider from the scope of the \textit{entire service chain} and propose our \textit{key observation} that when a vNF is overloaded, we can migrate \textit{other} vNFs on the SmartNIC away to \textit{release resources} for the overloaded vNF. To avoid introducing extra packet transmissions over PCIe, we choose to migrate vNFs on the \textit{border} of SmartNIC and CPU. As shown in Figure~\ref{fig:migrate-smartchain}, we migrate the \textit{Logger} to CPU to alleviate the \textit{Monitor} hot spot. However, selecting the right border vNFs for migration is challenging. Migrating too few vNFs may not effectively alleviate the hot spot, while migrating too many vNFs may waste CPU resource. To address this challenge, \textsf{PAM} carefully models SmartNIC and CPU resources and proposes an effective algorithm to find the most suitable vNFs for migration. Our evaluation shows that \textsf{PAM} could effectively alleviate the hot spot on SmartNIC and generate a service chain with 18\% lower latency compared with the naive solution. %reduce the long-term service chain latency by 20\% with the real world service chain in Figure~\ref{fig:migrate} comparing to the naive solution. 

\section{PAM Design}

In this section, we first introduce the resource constraints of SmartNIC and CPU. We then introduce how \textsf{PAM} identifies proper border elements on SmartNIC for migration to effectively alleviate the hot spots on SmartNIC without performance degradation due to extra packet transmissions.% and consistency ensured. 

To understand the resource constraints of the CPU and SmartNIC, we refer to~\cite{icc2018coco} and assume that \textit{the resource utilization of a vNF on both SmartNIC and CPU increases linearly with the vNF throughput}. % Thus we first measure the peak throughput capacity of each vNF on CPU and SmartNIC respectively. 
Suppose the throughput capacity of vNF $i$ on SmartNIC is $\theta_i^\mathcal{S}$ and the current throughput is $\theta_{cur}$, the ratio of consumed resource on SmartNIC is $\theta_{cur}/\theta_i^\mathcal{S}$. We measure and present the capacity of several vNFs in Table~\ref{tab:capacity}. 
We adopt the NF migration mechanism between SmartNIC and CPU introduced in~\cite{socc2017uno}. The network administrators can periodically query the load of SmartNIC and CPU and execute the \textsf{PAM} border vNF selection algorithm:

\begin{table}
\centering
\caption{Capacity of vNFs on the SmartNIC and CPU.}
\label{tab:capacity}
\vspace{-1\baselineskip}
\footnotesize
\begin{tabular}{|c|c|c|c|c|}
\hline
vNF $i$ & Firewall & Logger & Monitor & Load Balancer \\
\hline
$\theta_{i}^\mathcal{S}$ & 10 Gbps & 2 Gbps & 3.2 Gbps & $>$10 Gbps \\
\hline
$\theta_{i}^\mathcal{C}$ & 4 Gbps & 4 Gbps & 10 Gbps & 4 Gbps \\
\hline
\end{tabular}
\vspace{-1.5\baselineskip}
\end{table}

\noindent\textbf{Step 1: Border vNFs Identification.} We first find out the border vNFs of SmartNIC. We classify them into left border and right border vNFs, whose upstream or downstream vNF is placed on CPU. For example, the left border vNF in Figure~\ref{fig:migrate-origin} is \textit{Logger} and the right border vNF is \textit{Firewall}. Due to the several packet transmissions between SmartNIC and CPU, there may be multiple border vNFs in a service chain. We respectively denote them as set $\mathcal{B}_{L}$ and $\mathcal{B}_{R}$. \textit{Migrating border vNFs will not introduce new packet transmissions.} 

\noindent\textbf{Step 2: Migration vNF Selection.} To alleviate the overload with minimum number of vNF to migrate, we select the vNF $b_0$ from border vNFs with minimum capacity on SmartNIC:
\begin{equation}
\label{eq:select}
b_0=\mathop{\arg\min}_{b\in\mathcal{B}_{L}\cup\mathcal{B}_{R}} \theta_b^{\mathcal{S}}
\end{equation}

\noindent\textbf{Step 3: Overload Alleviation Check.} Meanwhile, we need to ensure \circled{1} migration will not cause new hot spots on CPU, and \circled{2} the overload of SmartNIC can be alleviated. For \circled{1}:%We denote the set of vNFs placed on CPU and SmartNIC as $\mathcal{V}_\mathcal{C}$ and $\mathcal{V}_\mathcal{S}$. 
% We first check the constraint \circled{1}:
\begin{equation}
\label{eq:mig-constr-1}
\vspace{-.2\baselineskip}
\sum_{i\in\{ vNFs~on~\mathcal{C}\}}\frac{\theta_{cur}}{\theta^\mathcal{C}_i}+\frac{\theta_{cur}}{\theta^{\mathcal{C}}_{b_0}}<1
\vspace{-.2\baselineskip}
\end{equation}
If Equation~\ref{eq:mig-constr-1} is not satisfied, which indicates migration will create new hot spots on CPU, we cannot migrate it to CPU. We remove $b_0$ from $\mathcal{B}_{L}$ or $\mathcal{B}_{R}$ and go back to Step 2. Otherwise, we can continue to check constraint \circled{2}:
\begin{equation}
\label{eq:mig-constr-2}
\sum_{i\in\{ vNFs~on~\mathcal{S}\},~i\neq b_0} \frac{\theta_{cur}}{\theta^\mathcal{S}_i} <1
\end{equation}
The algorithm terminates if Equation~\ref{eq:mig-constr-2} is satisfied. Otherwise, we migrate $b_0$ to CPU. If $b_0\in\mathcal{B}_{L}$, we remove it from $\mathcal{B}_{L}$ and add its downstream element into the set if the downstream element is also placed on SmartNIC. If $b_0\in\mathcal{B}_{R}$, we execute similar actions on its upstream element. We then go back to Step 2 to continue the loop. If both CPU and SmartNIC are overloaded, which rarely happens, the network operator must start another instance to alleviate the hot spot~\cite{sigcomm2014opennf}. % We can also drop packets at the ingress of the SmartNIC according to flow priority in advance to quickly alleviate the overload. 

%\noindent\textbf{Discussion.} Although \textsf{PAM} is based on the scenario of migrating the service chain between SmartNIC and CPU, the idea of \textsf{PAM} can also be used at the deployment of the service chain across multiple Virtual Machines or multiple servers.

\vspace{-.5\baselineskip}
\section{Preliminary Evaluation}

\begin{figure}
\centering
\vspace{-.5\baselineskip}
\subfigure[Latency]{
\includegraphics[scale=0.25]{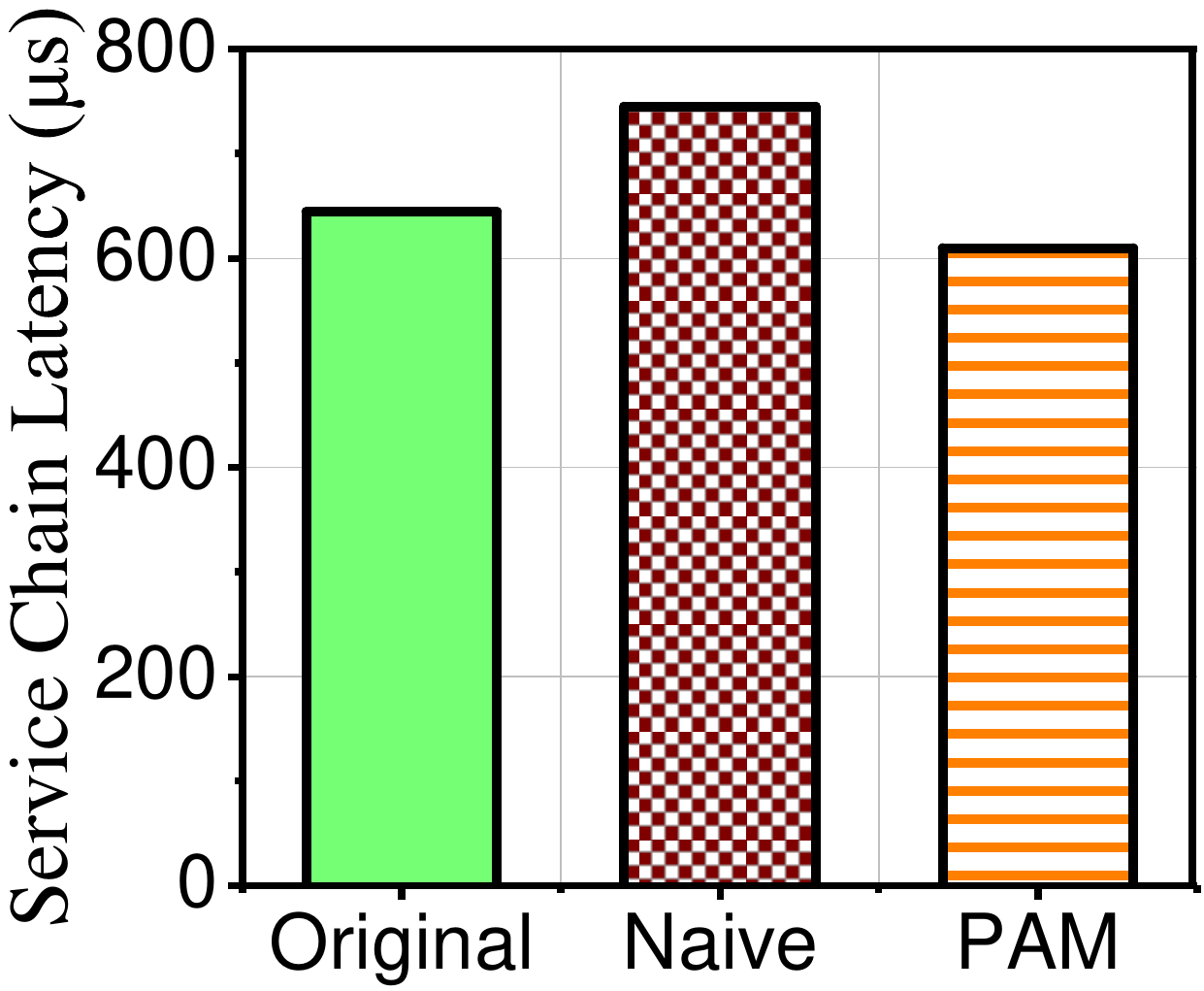}
\label{fig:eva-latency}
}
\subfigure[Throughput]{
\includegraphics[scale=0.25]{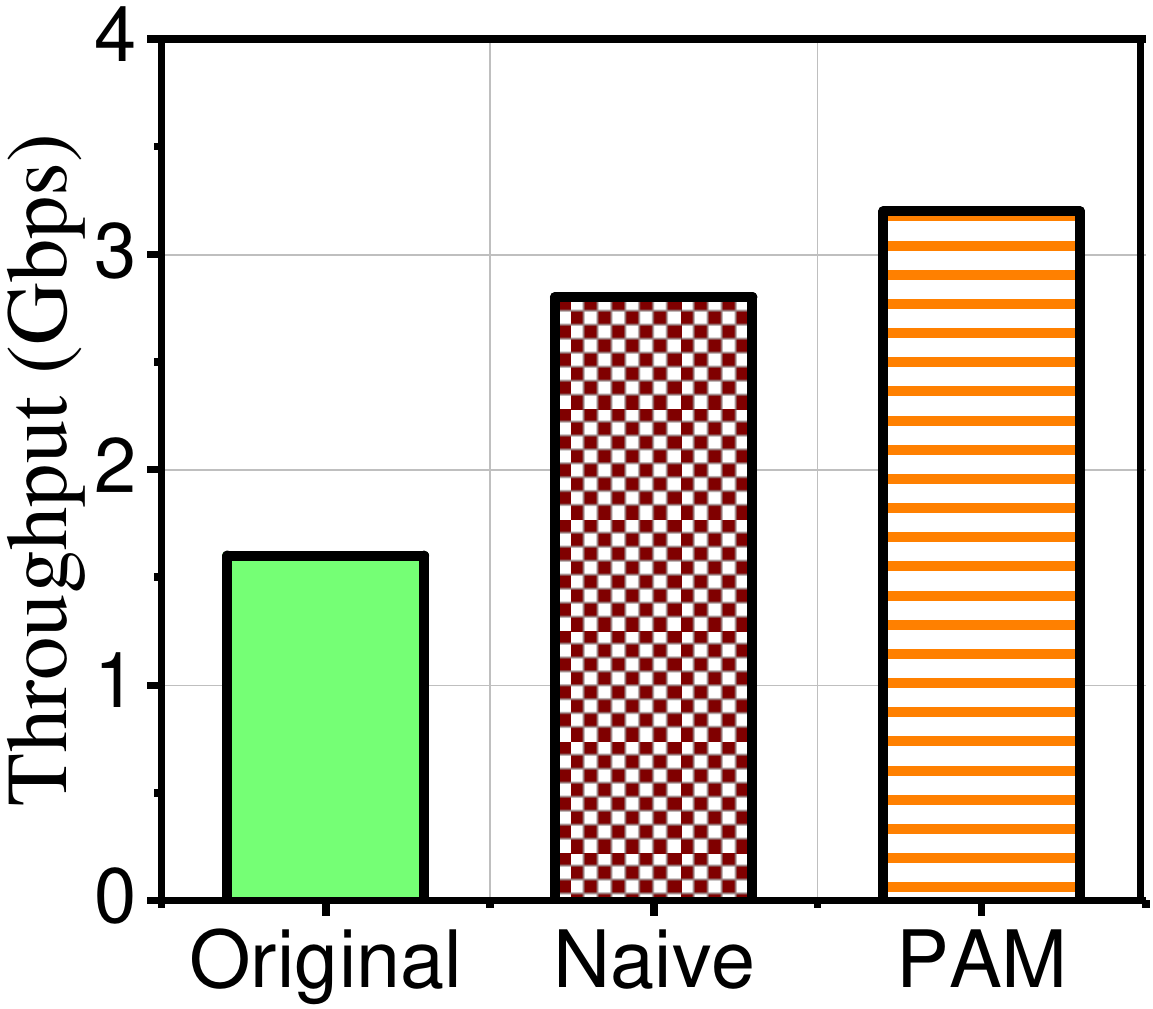}
\label{fig:eva-throughput}
}
\vspace{-1\baselineskip}
\caption{Comparison of the naive solution and \textsf{PAM}.}
\label{fig:eva}
\vspace{-1.5\baselineskip}
\end{figure}

We implement the service chain in Figure~\ref{fig:migrate} on a server equipped with one Netronome Agilio CX 2$\times$10GbE SmartNIC~\cite{netronome-smartnic-cx10gb}, two Intel Xeon E5-2620 v2 CPUs (2.10 GHz, 6 physical cores), and 128G RAM. For the naive algorithm, we pick the vNF on SmartNIC with minimal capacity $\theta_{NF}^\mathcal{S}$. We measure the service chain throughput and latency of different migration selection mechanisms in Figure~\ref{fig:migrate}. We vary the packet size from 64B to 1500B with a DPDK packet sender~\cite{dpdk} and present the average latency and throughput in Figure~\ref{fig:eva}.

\textsf{PAM} decreases the service chain latency by 18\% on average compared to the naive solution. The service chain latency with \textsf{PAM} is almost unchanged compared to the latency before migration because \textsf{PAM} does not introduce redundant packet transmissions. Meanwhile, the throughput of the service chain of \textsf{PAM} is improved a little since NFs may perform differently on SmartNIC and CPU.

\vspace{-.5\baselineskip}
\section{Conclusion and Future Work}

We have proposed a vNF selection scheme, \textsf{PAM}, which reduces the \textit{service chain latency} when alleviating hot spots on SmartNIC. We present our key novelty of pushing the \textit{border vNFs} aside to release resources for the \textit{bottleneck vNF}. Evaluation shows that \textsf{PAM} could alleviate the hot spot on SmartNIC and generate a service chain with 18\% lower latency compared with the naive solution. As our future work, we will analyze PCIe transmissions in detail, consider the difference of processing the same vNF on both devices, and extend \textsf{PAM} to work in FPGA-based SmartNICs.%set priorities for vNFs to enable more flexibility and customization for the scheme. 

\bibliographystyle{ACM-Reference-Format}
\bibliography{bibfile}
\end{document}